%% file: main.tex
\documentclass[conference]{IEEEtran}
\IEEEoverridecommandlockouts
\usepackage{amsmath,amssymb,amsfonts}
\usepackage{algorithmic}
\usepackage{graphicx}
\usepackage{textcomp}
\usepackage{xcolor}
\usepackage{subfig}
\usepackage{caption} 
\usepackage[nolist]{acronym}
\usepackage{setspace}
\usepackage{comment}
\def\BibTeX{{\rm B\kern-.05em{\sc i\kern-.025em b}\kern-.08em
    T\kern-.1667em\lower.7ex\hbox{E}\kern-.125emX}}
\usepackage{cite}
\usepackage[hyphens]{url}
\usepackage[hidelinks]{hyperref}
\usepackage{array}

\newcolumntype{C}{>{\centering\arraybackslash}m{1.28cm}}
\renewcommand{\baselinestretch}{0.96}
\begin{document}
\bstctlcite{IEEEexample:BSTcontrol}
\title{5G Campus Network Factory Floor Measurements with Varying Channel and QoS Flow Priorities}

\author{
	\IEEEauthorblockN{	Damir Hamidovic\IEEEauthorrefmark{1},
						Armin Hadziaganovic\IEEEauthorrefmark{1},
						Raheeb Muzaffar\IEEEauthorrefmark{1},
	                   Hans-Peter Bernhard\IEEEauthorrefmark{1}\IEEEauthorrefmark{2}
					}\\
	\IEEEauthorblockA{\IEEEauthorrefmark{1}Silicon Austria Labs GmbH, Linz, Austria}
	\IEEEauthorblockA{\IEEEauthorrefmark{2}Johannes Kepler University Linz, Austria}
	{firstname.lastname@silicon-austria.com}
}
\maketitle
\begin{singlespacing}
	\input{acronyms.tex}
\end{singlespacing}
\begin{abstract}
5G is considered a promising wireless communication technology to fulfill the high-demanding communication requirements of many Industry 4.0 applications. This work evaluates a 5G campus network for indoor factory floor scenarios using the latest commercially available 3GPP release-16 developments. The measurement campaign is conducted to obtain detailed coverage maps with reference signal received power, channel quality indicators, and downlink and uplink throughput (TP). Moreover, end-to-end delay measurements with varying channel conditions, 5G quality of service (QoS) priorities, and traffic loads were evaluated.  It was concluded that even without ultra-reliable low-latency features, the TP and latency performance could be controlled by configuring QoS parameters. The evaluations suggest scenarios where QoS allocation and retention priority levels can be used in order to ensure the required performance of a specific QoS flow within the 5G system.
\end{abstract}
\section{Introduction}

Many applications from industrial automation to health care, automotive, etc., require reliable low-latency communication preferably with wireless connectivity~\cite{9524600}. In that regard, the \ac{3GPP} promises novel opportunities for various industrial \ac{IoT} and smart manufacturing use cases that have stringent communication requirements~\cite{standard:service_req_3gpp, IEEEexample:5G_ACIA_5G_traffic_model, 9299391, Gangakhedkar2018use, 9652097}. More specifically, \ac{3GPP} specifications include 5G campus (non-public) network~\cite{9299391} that is meant to provide service to a specific organization over the geographic scope of a campus area, e.g., an indoor factory floor.

Evaluation of the 5G campus network has been performed in a few existing studies such as~\cite{9524600, 9977496, electronics11030412, electronics11111666, 8417704, 9935602, 9842810, fi13070180, 9502639, en14154444, 9238299}. However, only some reported valuable one-way \ac{OTA} measurements which are critical for the characterization of \ac{E2E} delays.  Others focused only on \ac{TP} or physical-layer measurements, e.g., \ac{RSRP} coverage whereby only some measurements were conducted for indoor industrial environments.

In this work, we provide performance evaluation of packet-level measurements, i.e., \ac{DL} and \ac{UL} \ac{TP} as well as one-way \ac{OTA} latency measurements in varying channel conditions and different \ac{QoS} parameters. Moreover, detailed coverage measurements of a factory floor environment, including various 5G parameters such as \ac{CQI}, \ac{MCS}, \ac{RSRP}, and \ac{TP} coverage maps, are presented. On one hand, the detailed coverage maps with more than 100 measurement points could serve as a guideline for \acp{RRU} placement and planning of a 5G deployment in factory floor environments. On the other hand, the packet-level measurements can help decide which parameters should be considered more specifically in order to fulfill the requirements of certain industrial use cases, as defined by \ac{3GPP} \cite{standard:service_req_3gpp}. Some of 5G \ac{QoS} parameters that can be used for guaranteed or non-guaranteed traffic control are \ac{5QI}, \ac{GFBR}, \ac{MFBR} or \ac{AMBR}~\cite{standard:ARP}. Certainly, \ac{URLLC} can provide new possibilities for mission-critical or high-demanding industrial applications, however, \ac{URLLC} features are still not commercially available.

In this work, we show through an example how the 5G \ac{QoS} \ac{ARP} parameter influences one-way \ac{DL} and \ac{UL} \ac{OTA} latency and \ac{TP} performance. We also show that a specific QoS flow can be prioritized by allocating a higher number of resources. Similarly, we present how the 5G \ac{QoS} parameters can be configured to increase reliability for specific industrial applications.

The remainder of the paper is organized as follows.  Section~\ref{sec:meas_setup} presents the detailed measurement setup, section~\ref{sec:meas_results} explains the measurement results in three subsections: (i) factory floor heatmaps of various measured parameters, (ii) TP performance results, and (iii) one-way latency results, each with varying channel conditions, QoS flow priorities, and traffic load. Finally, section~\ref{sec:conclusion} emphasizes the most important results to conclude the paper.

\section{Measurement setup of the 5G testbed}
\label{sec:meas_setup}
\begin{figure}[t!]
	\centering
	\includegraphics[width=0.45\textwidth]{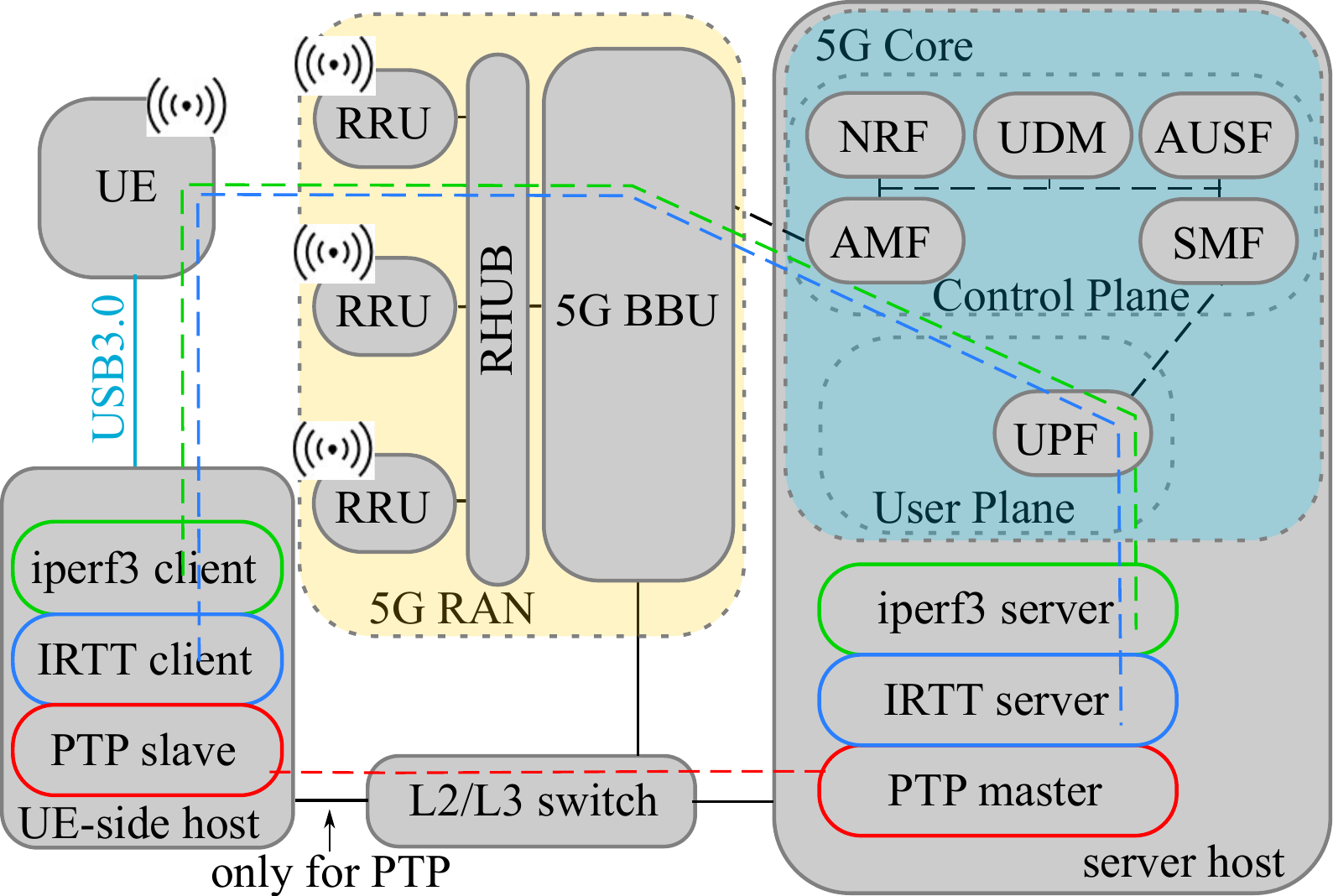}
	\caption{Measurement setup of the 5G testbed.}
	\label{fig:setup}
        \vspace{-3mm}
\end{figure}
The measurement setup is reflected in Fig.$\,$\ref{fig:setup}, where the solid black lines represent wired Ethernet connections, while the dashed lines represent logical connections of the setup. The \ac{5GS} consists of the 5G core, 5G \ac{RAN}, and \acp{UE}. This \ac{SA} 5G testbed setup uses 80$\,$MHz of \ac{BW} at 3.41-3.49$\,$GHz in band n78, full \ac{RB} allocation, \ac{SCS} of 30$\,$kHz and a \ac{TDD} configuration 4:1 (DDDDU).

The 5G core is Open5GCore\footnote{[Online]: \href{https://www.open5gcore.org/}{https://www.open5gcore.org/}} and is deployed as bare-metal. In such a deployment, the 5G core components including the \ac{AMF}, \ac{SMF}, \ac{NRF}, \ac{UDM}, \ac{AUSF}, and \ac{UPF} are running directly (without virtualization) on a host machine. The 5G core host is a Dell EMC PowerEdge R740XD server machine with Ubuntu 20.04 LTS \ac{OS}.

The \ac{RAN} part consists of a Huawei \ac{BBU} 5900, three \acp{RRU} 5973, each with 4 \ac{TX} and 4 \ac{RX} antennas with maximum TX power set to 24$\,$dBm, and a \ac{RHUB} 5963e. Quectel RM502Q-GL\footnote{[Online]: \href{https://www.quectel.com/product/5g-rm50xq-series}{https://www.quectel.com/product/5g-rm50xq-series}} 5G modules with Qualcomm Snapdragon X55 5G modem, supporting 4$\times$4$\,$\ac{MIMO} in DL and 2$\times$2$\,$\ac{MIMO} in UL, were used as the \acp{UE}. Each \ac{UE} is connected to its host machine (e.g., a Raspberry Pi 4 module with Ubuntu OS) via USB 3.0 interface. The 5G core and the \ac{RAN} are connected via a Netgear switch.

The measurements were performed using three sever-client applications executed on the server-host and UE-side hosts, respectively. These applications include the \textit{\ac{iperf}3} for real-time \ac{DL} and \ac{UL} \ac{TP} measurements and for the traffic generation, \textit{\ac{IRTT}} for one-way \ac{OTA} latency measurements and \textit{\ac{ptp4l}} for synchronization between the server host and the UE-side host \acp{NIC}. In order to synchronize the \textit{\ac{IRTT}}\footnote{[Online]: \href{https://github.com/heistp/irtt}{https://github.com/heistp/irtt}} server and client applications with their \acp{NIC}, the \textit{phy2sys} tool was used. In our setup, hardware timestamps were used for precise \ac{PTP} synchronization. Therefore, to synchronize UE hosts using \ac{PTP}, Ethernet interface between the UE-side host and the switch was used to exchange \ac{PTP} packets only, as is depicted in Fig.$\,$\ref{fig:setup}. The \ac{PTP} synchronization precision was under 2$~\mu$s for the measurement duration lasting for an hour.
\section{Measurement results}
\label{sec:meas_results}
\subsection{Factory floor performance heatmaps}
\begin{figure}[t!]
	\centering
	\subfloat[PDSCH CQI (range 0-15).\label{fig:CQI}]{%
		\includegraphics[width=0.44\textwidth]{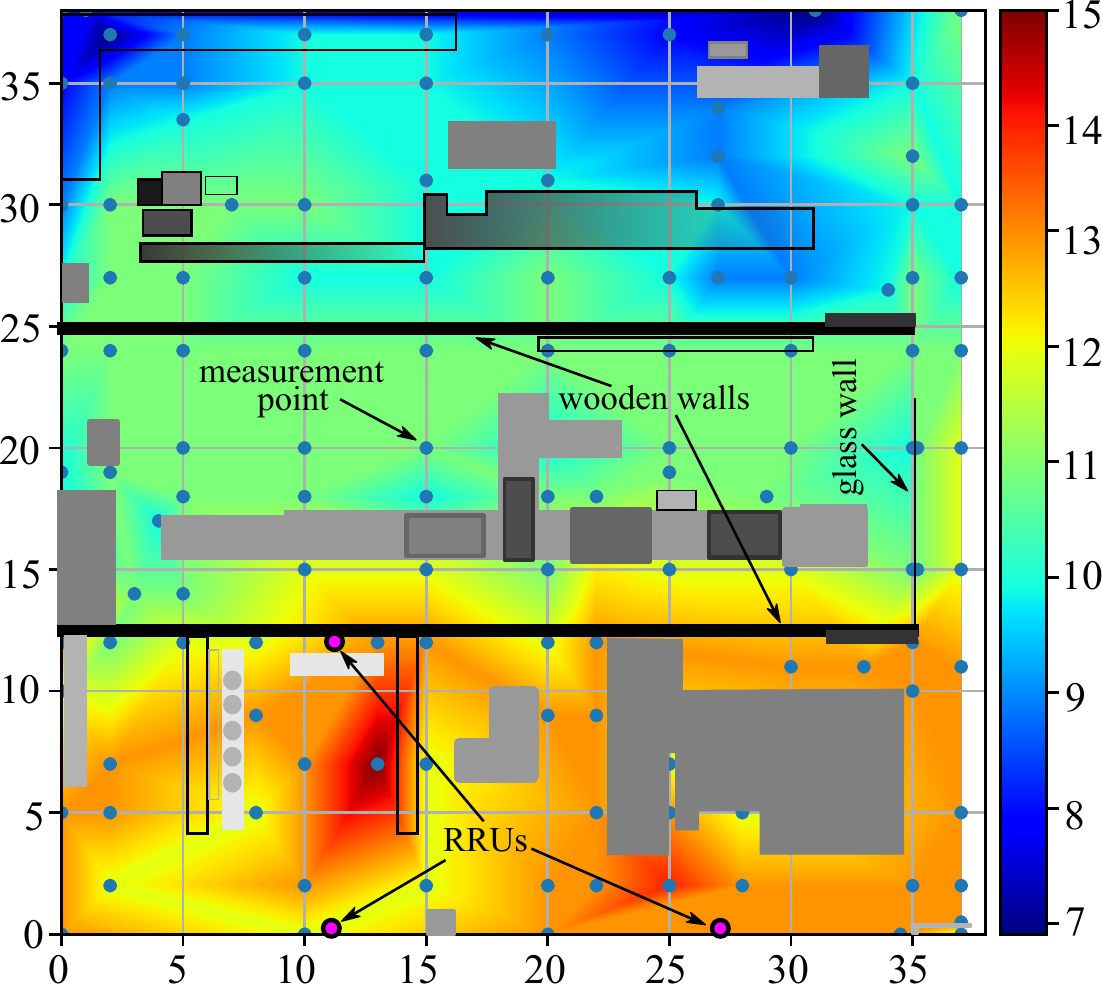}
	}\\
	\subfloat[PDSCH MCS (range 0-28, 256-QAM supported).\label{fig:MCS}]{%
		\includegraphics[width=0.44\textwidth]{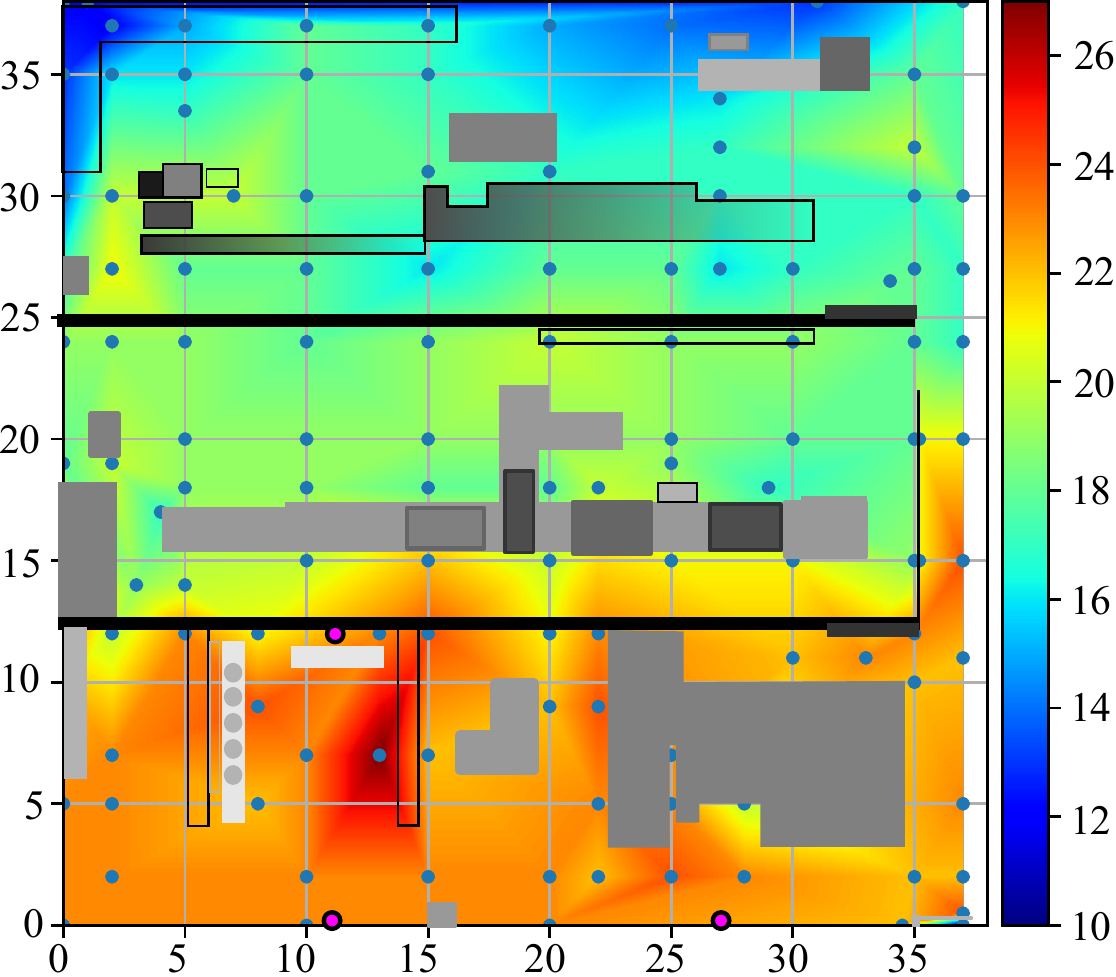}
	}
	\caption{CSI heatmaps of the factory floor.}
	\label{fig:heatmaps_CSI}
        \vspace{-3mm}
\end{figure}

\begin{figure}[t!]
	\centering
	\subfloat[DL Throughput (Mbps).\label{fig:TP_DL}]{%
		\includegraphics[width=0.44\textwidth]{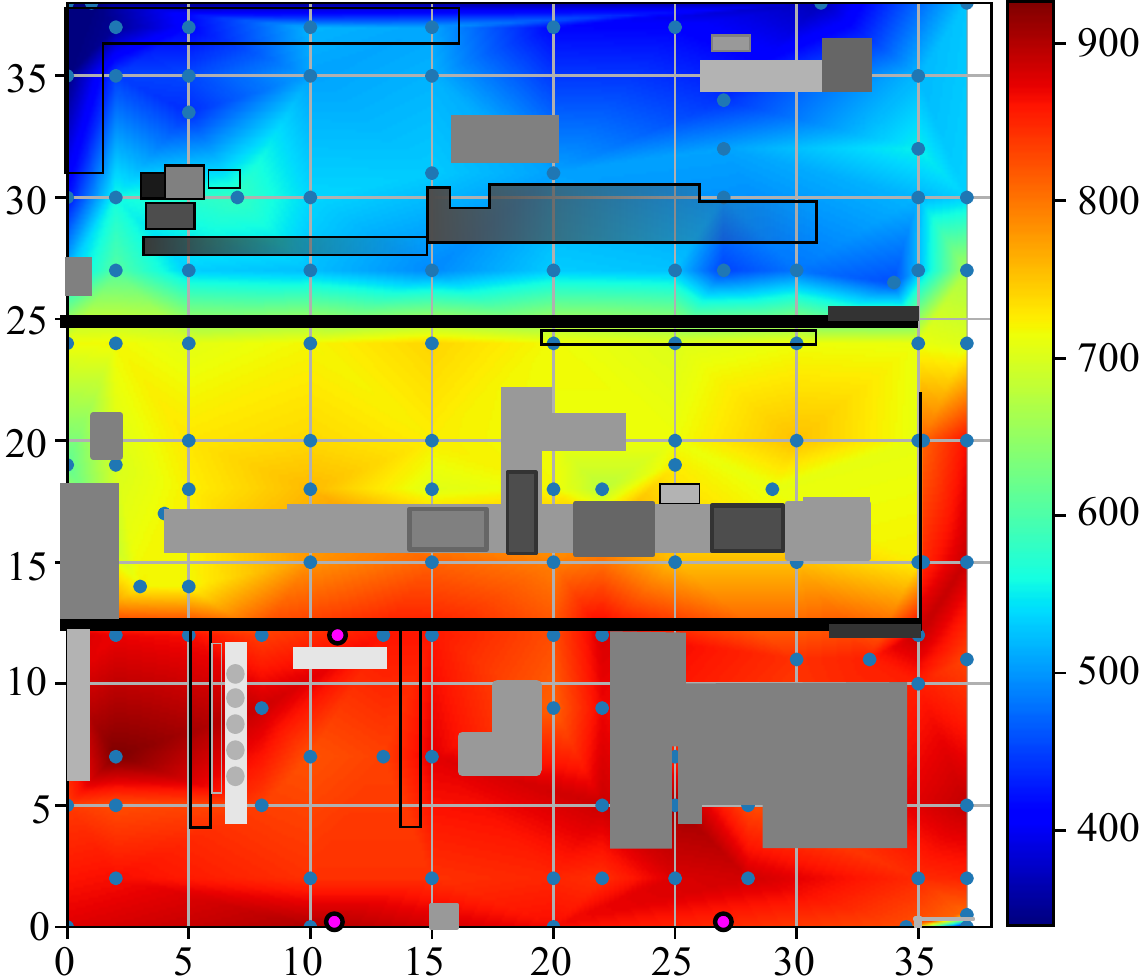}
	}\\
	\subfloat[UL Throughput (Mbps).\label{fig:TP_UL}]{%
		\includegraphics[width=0.44\textwidth]{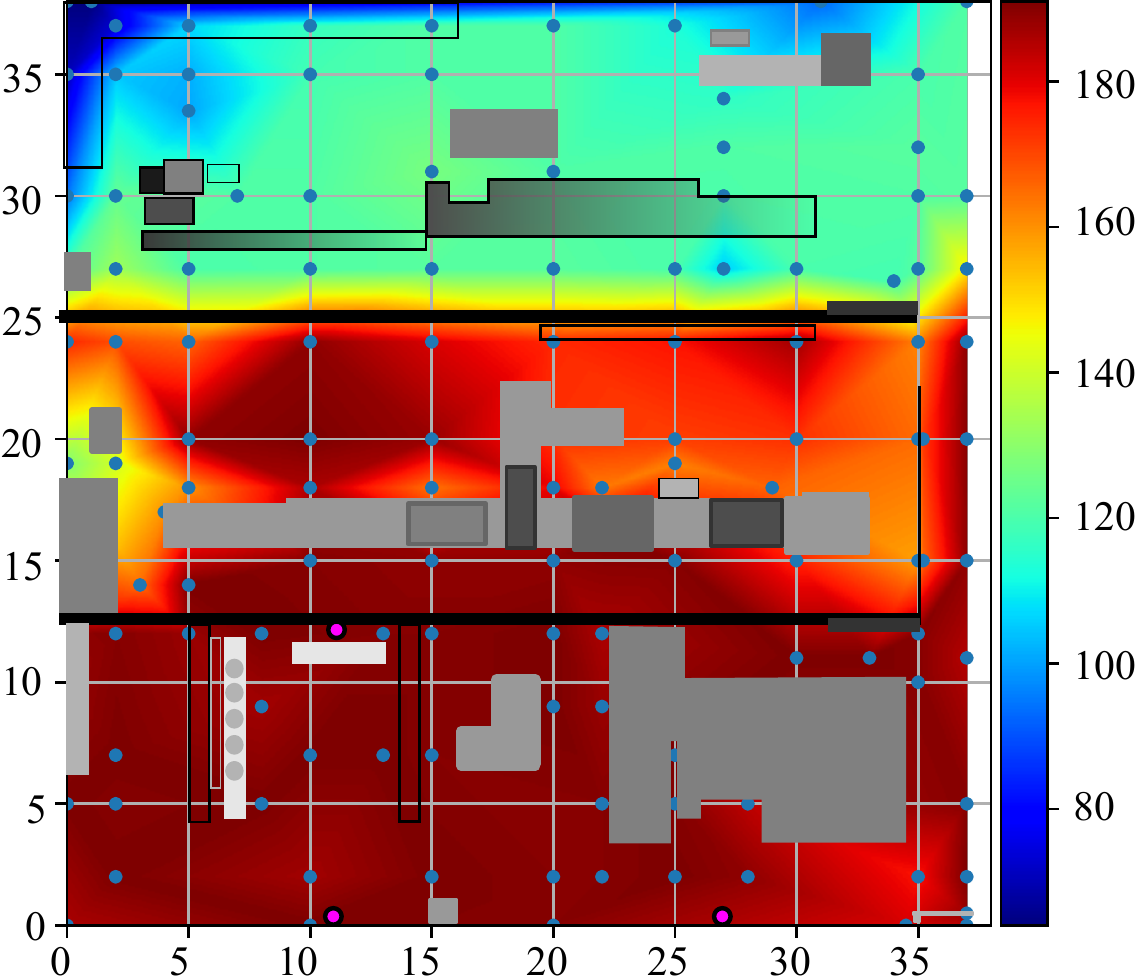}
	}
	\caption{TP heatmaps of the factory floor.}
	\label{fig:heatmaps_TP}
        \vspace{-3mm}
\end{figure}

\begin{figure}[t!]
	\centering
	\includegraphics[width=0.44\textwidth]{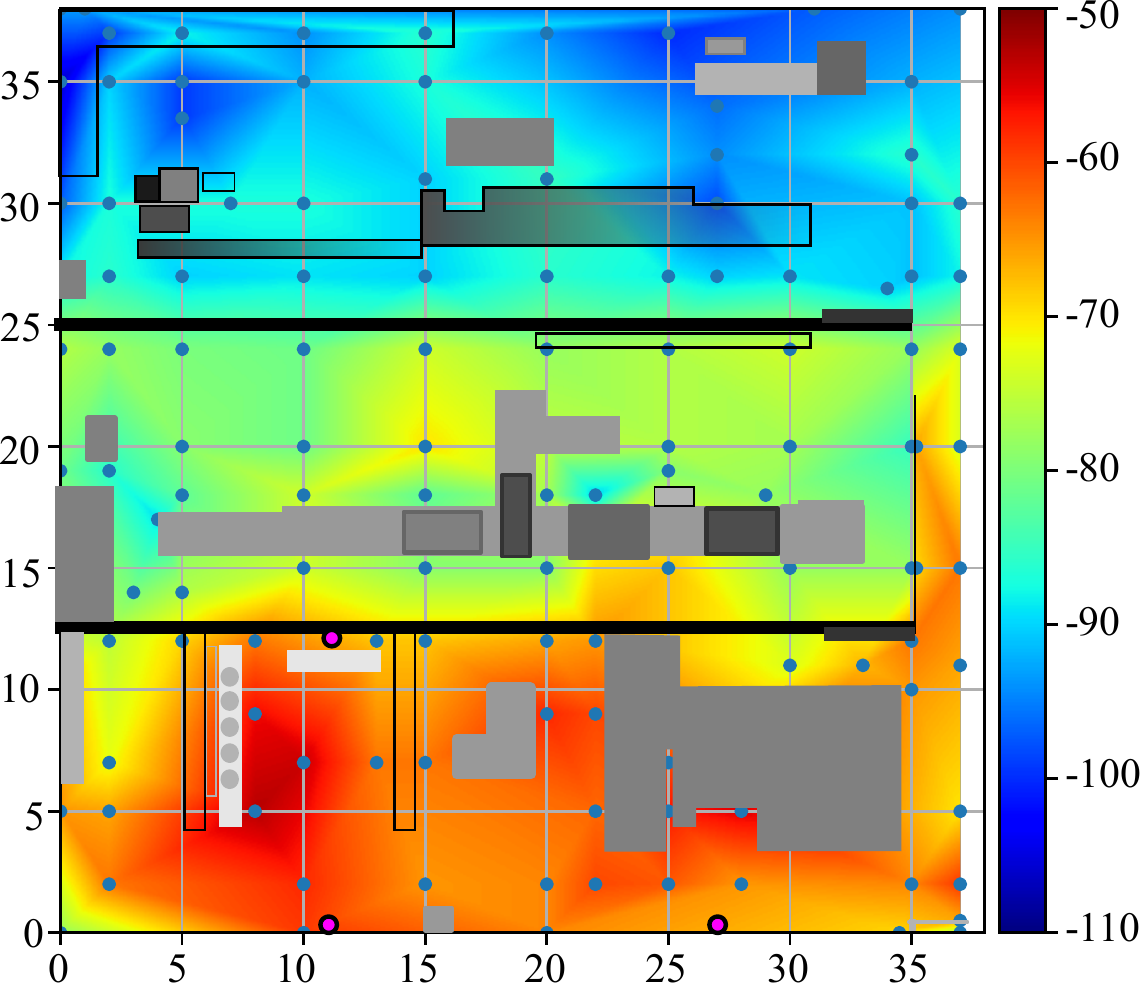}
	\caption{RSRP (dBm) heatmap of the factory floor.}
	\label{fig:RSRP_PRX}
        \vspace{-3mm}
\end{figure}

In this section, we present coverage measurement results of the \ac{5GS} performed on the \ac{LIT} factory floor with dimensions 38$\,$m$\,\times\,$38$\,$m, located in the \ac{JKU} campus. The obstacles on the factory floor are predominantly steel machines with different heights, as represented in Fig.$\,$\ref{fig:heatmaps_CSI}-\ref{fig:RSRP_PRX} in coarse grayscale. The white color represents 0$\,$m and the black color represents a height of 10$\,$m. Note that horizontal and vertical axes in Fig.$\,$\ref{fig:heatmaps_CSI}-\ref{fig:RSRP_PRX} represent distance in meters. The \ac{RSRP} and \ac{CSI}, i.e., \ac{CQI} and \ac{MCS}, were collected at the \ac{UE} side. Each parameter at every measurement point is averaged over five consecutive measurements.

Fig.$\,$\ref{fig:CQI} and \ref{fig:MCS} show heatmaps of the \ac{CSI} parameters, \ac{CQI} (range: 0-15) and \ac{PDSCH} \ac{MCS} (range: 0-28), measured during an \textit{iperf3} test for maximum \ac{DL} \ac{TP}. Note that 256-\ac{QAM} scheme is supported in both UL and DL. Fig.$\,$\ref{fig:TP_DL} and \ref{fig:TP_UL} show heatmaps of the \ac{DL} and \ac{UL} \ac{TP}, measured and averaged over a duration of 5$\,$s of an \textit{iperf3} test for maximum \ac{DL} and \ac{UL} traffic load, respectively. Fig.$\,$\ref{fig:RSRP_PRX} shows the heatmap of the \ac{RSRP} values (range: from -140 to -44$\,$dBm). 

From the presented heatmaps, it can be seen that the entire factory floor area has good coverage with 5G signals, i.e., more than 450$\,$Mbps of \ac{DL} \ac{TP} and more than 120$\,$Mbps of \ac{UL} \ac{TP} is available in the entire area, with the exception of very few measurement points. In the primary area, i.e., from coordinate points (0,0) to (38,12), excellent \ac{TP} performance with a minimum of 800$\,$Mbps of DL TP and 180$\,$Mbps of UL TP was recorded. However, the measurement point (37,0) is an exception due to the strong blockage of power-supply installations. It can be noted that the wooden walls, separating the factory floor in three areas, also attenuate the signal. However, high signal power and good \ac{TP} performance are available even beyond the wooden walls. It can be noted that the height of obstacles, e.g., at points (15,18), (22,18), or (29,18), shows a noticeable impact on the \ac{TP} performance due to signal blockage. It is also important to note that significant signal attenuation is caused by the glass wall positioned at (35,12)-(35,22). Moreover, the distance from \acp{RRU} shows a significant impact on the \ac{RSRP} and the \ac{TP} due to path loss.
\subsection{DL and UL TP measurements with varying channel conditions and QoS priorities}

In this section, the TP performance in different channel quality scenarios is characterized and the impact of the 5G \ac{QoS} \ac{ARP} parameter \cite{standard:ARP} is shown. The ARP priority level defines the relative importance of a QoS flow, where the range of priority levels is 1 to 15, with 1 being the highest priority. Priority weight $w_\text{i}$ defines the relative amount of resources $R_\text{i}$, that can be allocated to a \lq QoS $\text{flow\rq}_\text{i}$ and can be formulated in a simple form as 
\begin{equation}
	R_\text{i}=\frac{w_{\text{i}}}{\sum_{n=1}^{N}w_\text{n}},
	\label{eq:w}
\end{equation}
where $N$ is the number of competing QoS flows. However, a QoS flow can get more than $R_\text{i}$ resources, if available, i.e., if no other QoS flows are competing for the same resources. The weights $w$ are configured in the BBU, while the priority level for a specific QoS flow is defined in the 5G core, specifically in the \ac{UDM} database. Note that the weights $w$ can be controlled independently for DL and UL, however, in our scenarios, the same weights were used for DL and UL. We measured the TP performance in 5 different scenarios:
\begin{itemize}
    \item S1: Equal QoS parameters with good channel conditions.
    \item S2: Equal QoS parameters. UE1 has poor, while UE2 and UE3 have good channel conditions. 
    \item S3: Different QoS with good channel conditions.
    \item S4: Different QoS. UE1 has poor channel conditions, while UE2 and UE3 have good channel conditions.
    \item S5: Different QoS with good channel conditions. UE1 requests limited 300$\,$Mbps in DL or 30$\,$Mbps in UL, while UE2 and UE3 request maximum traffic load.
\end{itemize}
\begin{figure}[t!]
	\centering
	\includegraphics[width=0.45\textwidth]{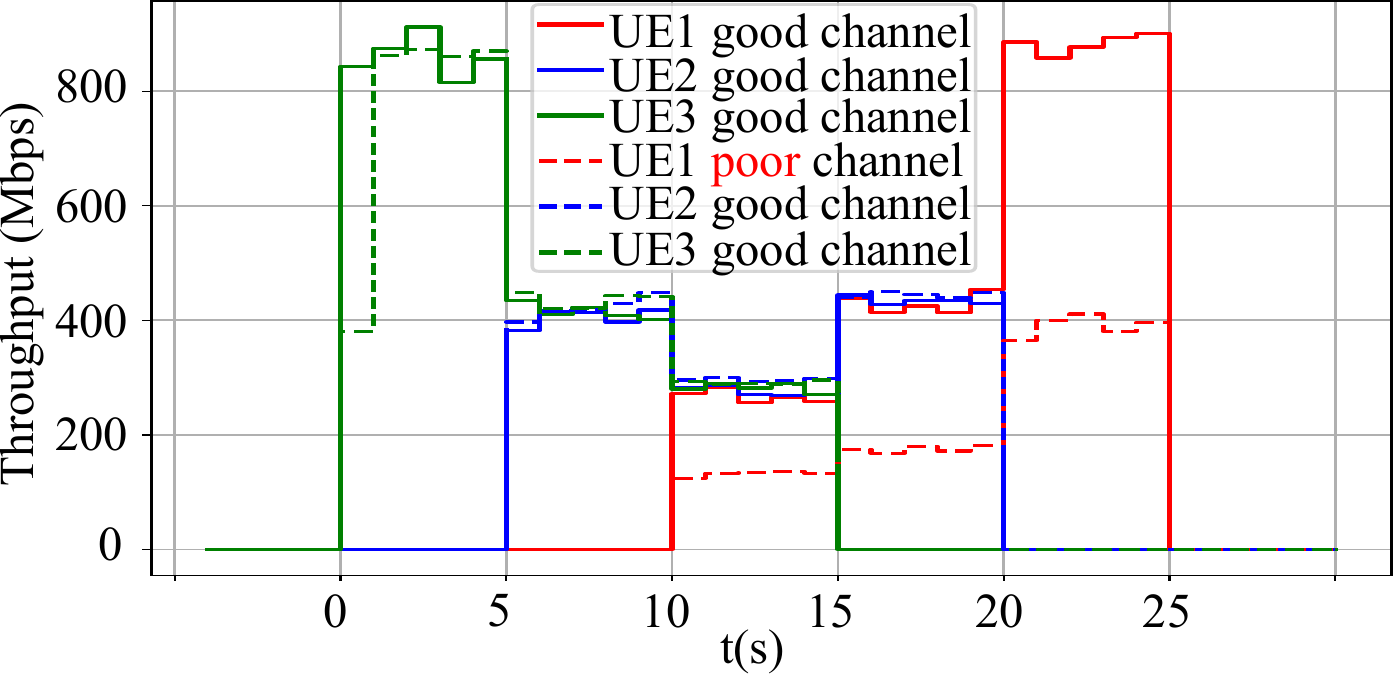}
	\caption{Impact of channel conditions on the DL TP by comparing scenarios S1 (solid lines) and S2 (dashed lines).}
	\label{fig:UC1_UC2}
        \vspace{-3mm}
\end{figure}
\begin{figure}[t!]
	\centering
	\includegraphics[width=0.45\textwidth]{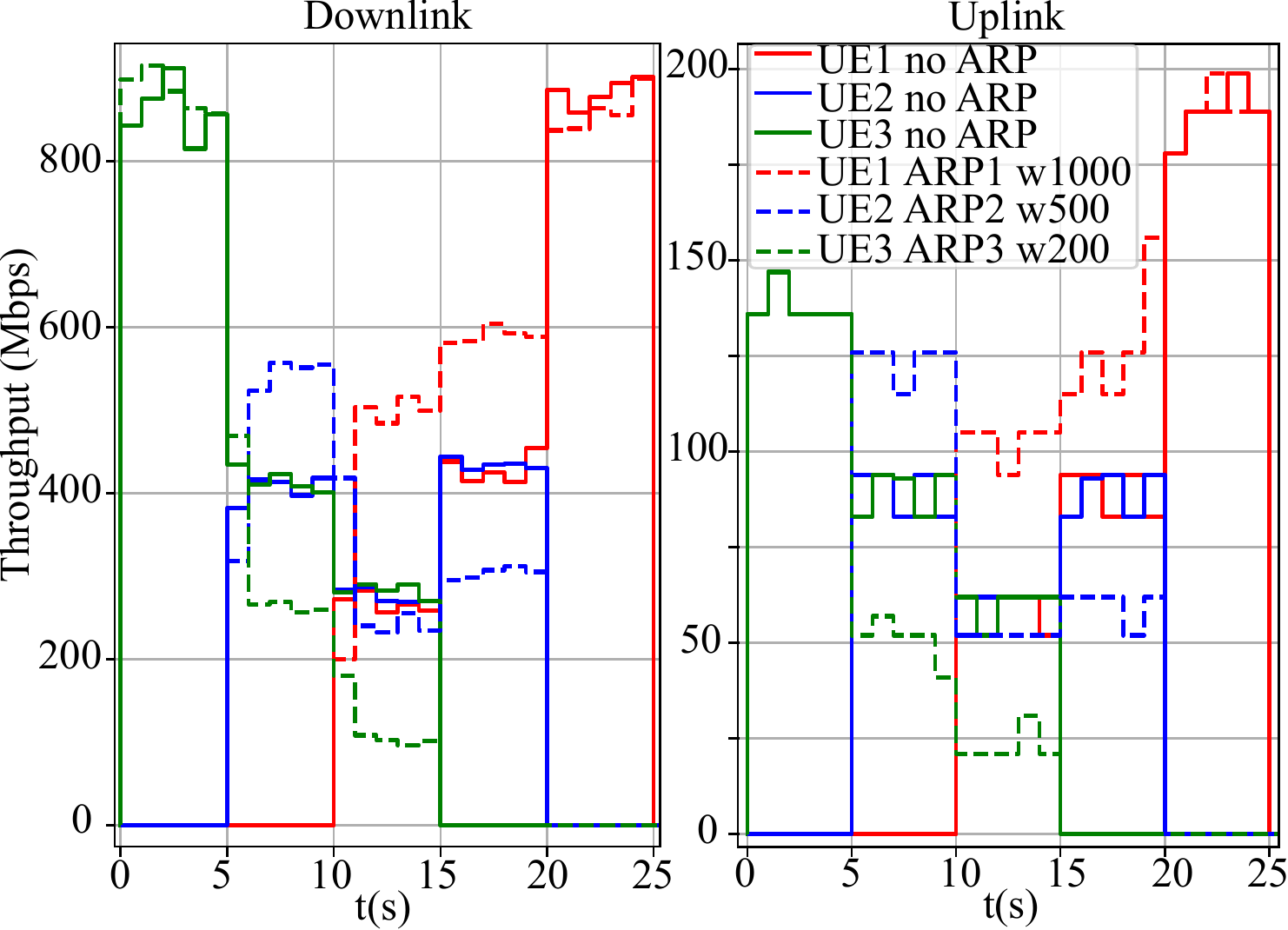}
	\caption{Impact of the QoS ARP level on the DL and UL TP by comparing scenarios S1 (solid lines) and S3 (dashed lines).}
	\label{fig:UC1_UC3}
        \vspace{-3mm}
\end{figure}
Different QoS means that UE1 has ARP priority level 1, UE2 2, and UE3 3, with weights $w_\text{UE1}$=1000, $w_\text{UE2}$=500, and $w_\text{UE3}$=200. From (\ref{eq:w}), $R_\text{UE1}$=1000/1700=0.59, $R_\text{UE2}$=500/1700=0.29 and $R_\text{UE2}$=200/1700=0.12.

Each UE in scenarios S1-S4 requests maximum traffic load. The \ac{TCP} traffic load for each UE, in scenario S1-S5, was generated using \textit{iperf3} tool in different time-frames as follows: UE1 (10-25)$\,$s, UE2 (5-20)$\,$s, UE3 (0-15)$\,$s. The reason for UEs being active in different time frames is to show the TP capacity distribution among UEs in scenarios when only one UE is active, two UEs are active and all three UEs are active. The TP performance with poor channel conditions was measured at point (2,37) of the factory floor, where measurement points are presented on heatmaps (see Fig.$\,$\ref{fig:CQI}, where it can be seen that the \ac{CQI} value at this point is 7). Note that the performance among only three UEs has been analyzed. However, each UE can represent a group of UEs with the same channel conditions and QoS flows, which share the same radio resources, in order to estimate the TP performance with a higher number of UEs.

In Fig.$\,$\ref{fig:UC1_UC2}, the scenario S1 (solid lines) is compared with the S2 (dashed lines). It can be seen that in S1 the available TP capacity is shared equally among active UEs. 
From time 10 to 25$\,$s, it can be seen that the DL TP performance of UE1 has decreased significantly in S2, due to poor channel conditions of UE1. UE1 achieves lower TP with the same amount of resources, due to a lower modulation scheme used.

In Fig.$\,$\ref{fig:UC1_UC3}, scenarios S1 (solid lines) and S3 (dashed lines) are compared, where the impact of the QoS ARP parameter on the TP performance, with all UEs with good channel conditions, is presented. It can be seen from time 5 to 20$\,$s, that both UL and DL TP performance in S3 is shared among active UEs equally to the share of priority weights $w$ since all UEs competing for the same resources have the same channel conditions.
\begin{figure}[t!]
	\centering
	\includegraphics[width=0.45\textwidth]{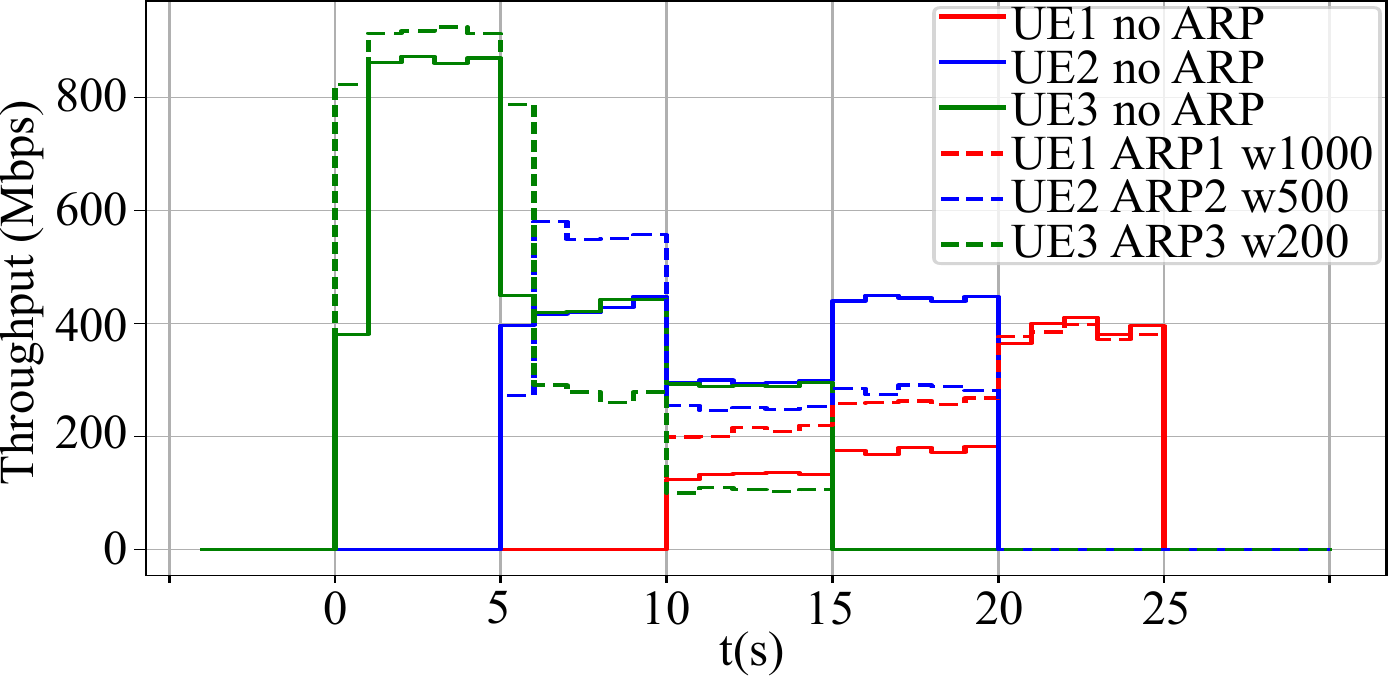}
	\caption{Impact of the QoS ARP level on the DL TP by comparing scenarios S2 (solid lines) and S4 (dashed lines).}
	\label{fig:UC2_UC4}
        \vspace{-3mm}
\end{figure}
\begin{figure}[t!]
	\centering
	\includegraphics[width=0.45\textwidth]{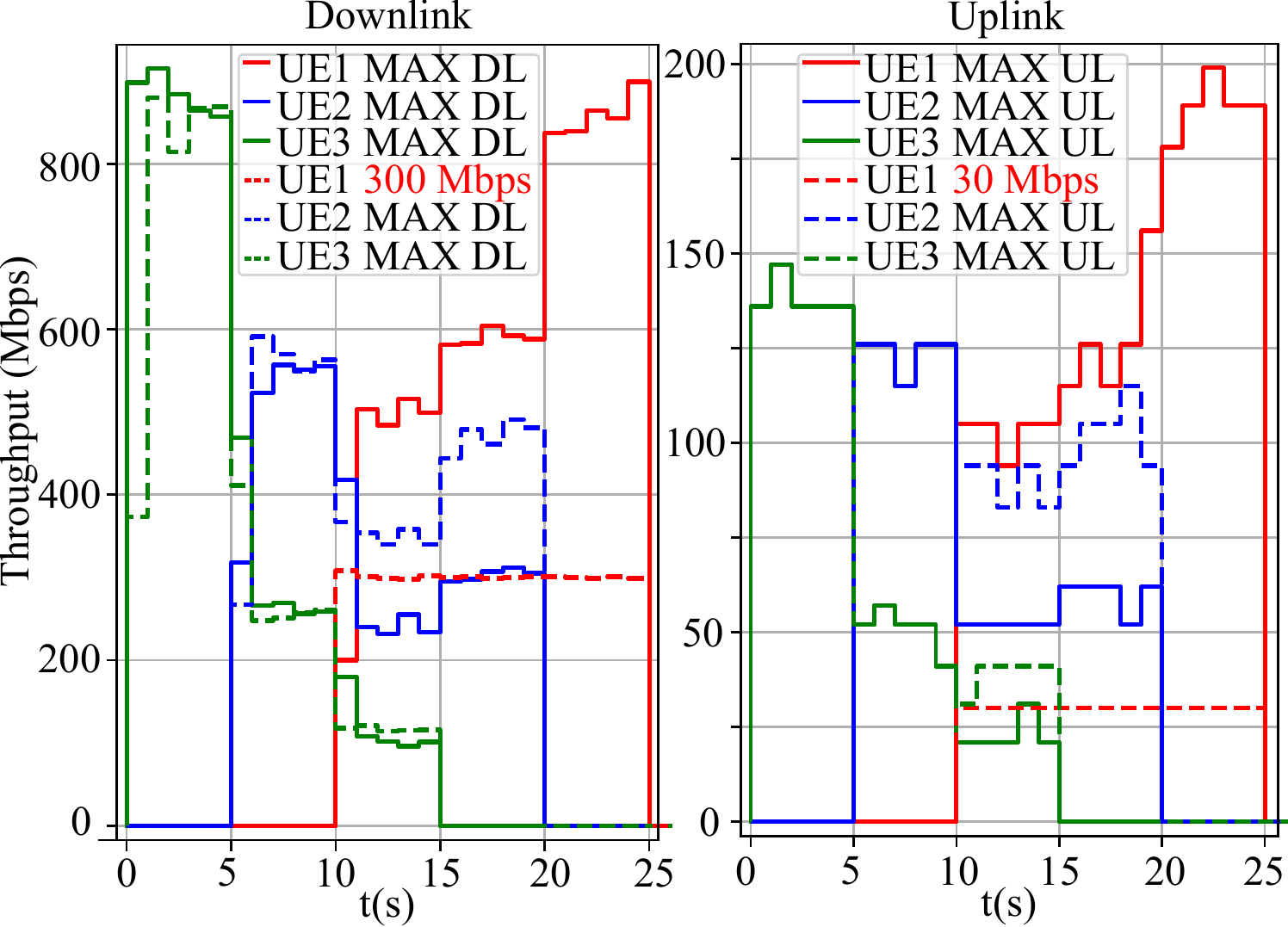}
	\caption{Impact of traffic load on the DL and UL TP by comparing scenarios S3 (solid lines) and S5 (dashed lines).}
	\label{fig:UC3_UC5}
        \vspace{-3mm}
\end{figure}

The TP performance of low-priority UEs does not have to be limited in a fixed manner in order to secure the needed resources for higher-priority UEs or QoS flows. Instead, we showed that the resources can be dynamically allocated with prioritization for higher-priority QoS flows for better utilization of the limited resources.

Similar to Fig.$\,$\ref{fig:UC1_UC3}, Fig.$\,$\ref{fig:UC2_UC4} compares scenarios S2 (solid lines) and S4 (dashed lines) to show the impact of QoS ARP on the DL TP performance of UEs. Here the UE1 is under poor channel conditions. It can be seen that if the high-priority UE1 has poor channel conditions, the QoS ARP could ensure that it gets more of the available resources (proportional to the ARP weights $w$) in order to improve the TP performance, which could be critical, especially under poor channel conditions.

However, it is important to control the requested resources of the high-priority UEs or QoS flows to maximize the utilization of resources such that lower-priority UEs can be serviced. The aforementioned scenario S5 (dashed lines) is compared with S3 (solid lines), as presented in Fig.$\,$\ref{fig:UC3_UC5}. It can be seen that even though the UE1 has the highest priority, by limiting the requested DL or UL TP performance to only a needed level, such as 300$\,$Mbps for DL and 30$\,$Mbps for UL, low priority UEs can utilize the available resources. We also noticed that UE3 decreased its maximum UL TP performance compared to UE1 even when the channel and QoS conditions were similar (compare time frames 0 to 5$\,$s and 20 to 25$\,$s in Fig.$\,$\ref{fig:UC1_UC3} and Fig.$\,$\ref{fig:UC3_UC5} for UL).

Based on the measurement results, it can be concluded that the resource share of a QoS flow can be controlled in a deterministic manner with the QoS ARP level in varying channel conditions and traffic loads. As an example, the TP performance is well suited for an industrial use case \lq Small-scale deployment scenario\rq, defined by 5G \ac{ACIA} in Table 12~\cite{IEEEexample:5G_ACIA_5G_traffic_model}, where 200 UEs running different applications require a total of 105$\,$Mbps for UL and 300$\,$Mbps for DL TP.

\subsection{One-way OTA UL and DL latency measurements with varying channel conditions and QoS priorities}
\begin{figure}[t!]
	\centering
	\subfloat[CCDF.\label{fig:CCDF_good_bad}]{%
		\includegraphics[width=0.48\textwidth]{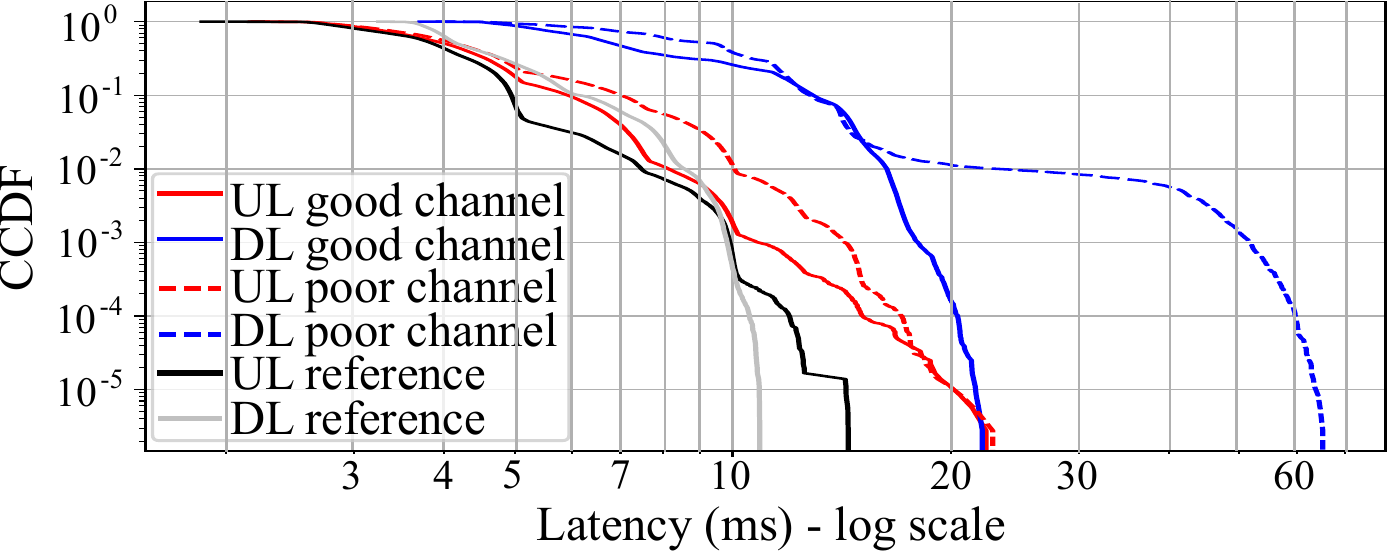}
	}\\
	\subfloat[PDF.\label{fig:PDF_good_bad}]{%
		\includegraphics[width=0.48\textwidth]{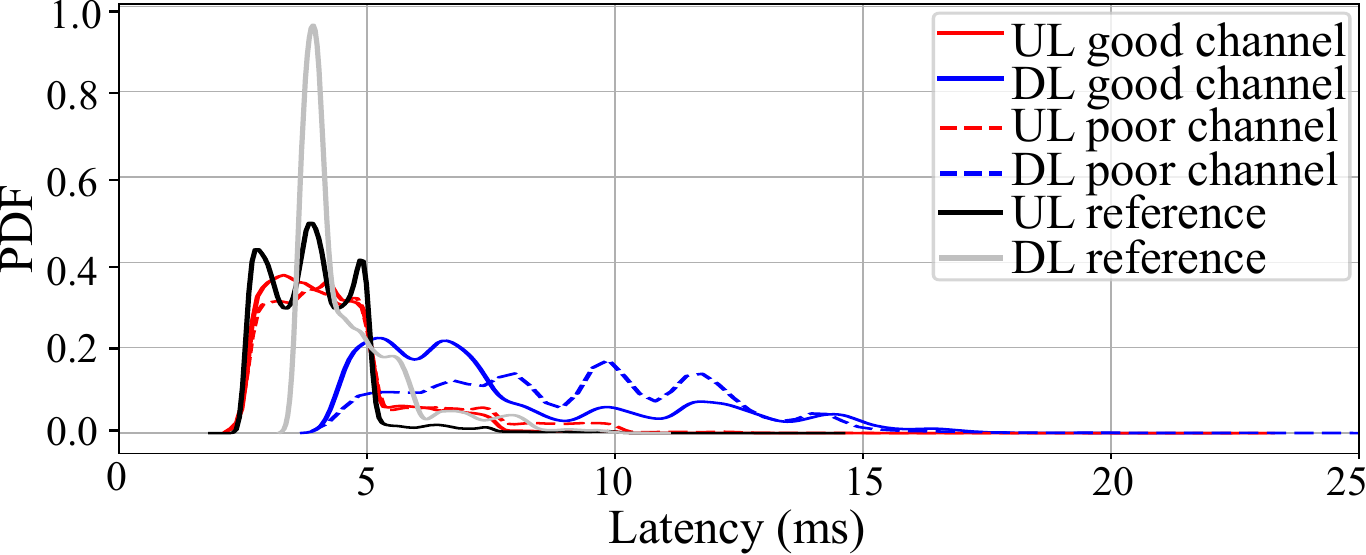}
	}
	\caption{Impact of the channel conditions on the UL and DL one-way \ac{OTA} latency performance of a UE with additional DL traffic at 450$\,$Mbps.  The reference scenario is the latency measurement of a single UE without additional traffic load.}
	\label{fig:latency_good_bad}
        \vspace{-3mm}
\end{figure}
\begin{figure}[t!]
	\centering
	\subfloat[CCDF.\label{fig:CCDF_2_UEs_QoS_no_QoS}]{%
		\includegraphics[width=0.48\textwidth]{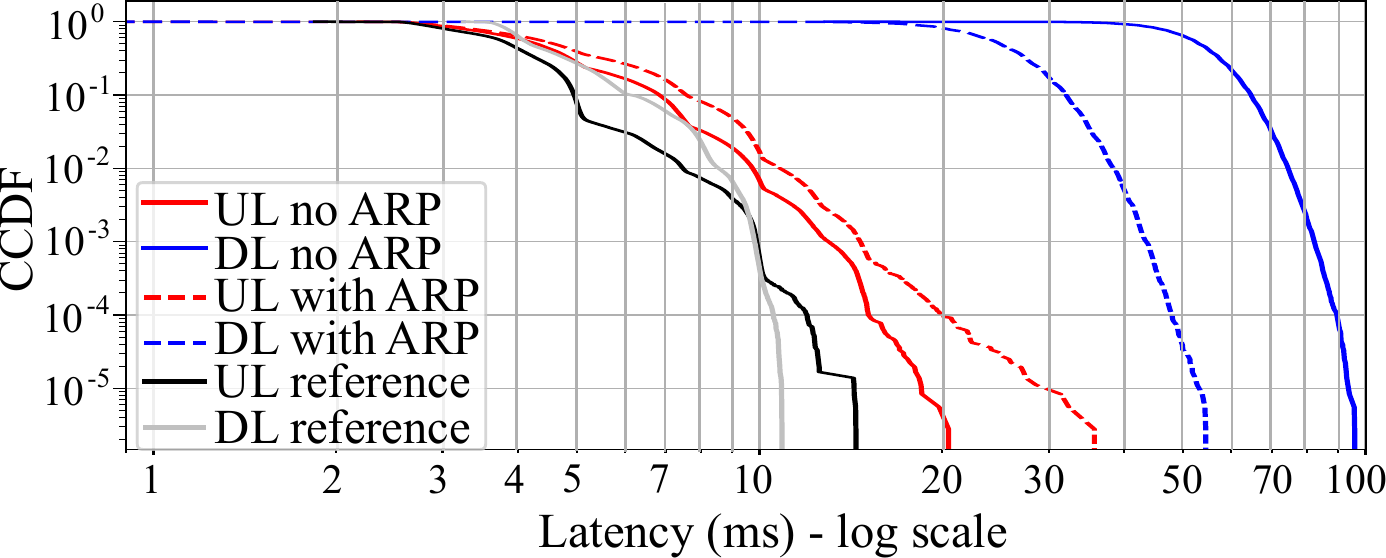}
	}\\
	\subfloat[PDF.\label{fig:PDF_2_UEs_QoS_no_QoS}]{%
		\includegraphics[width=0.48\textwidth]{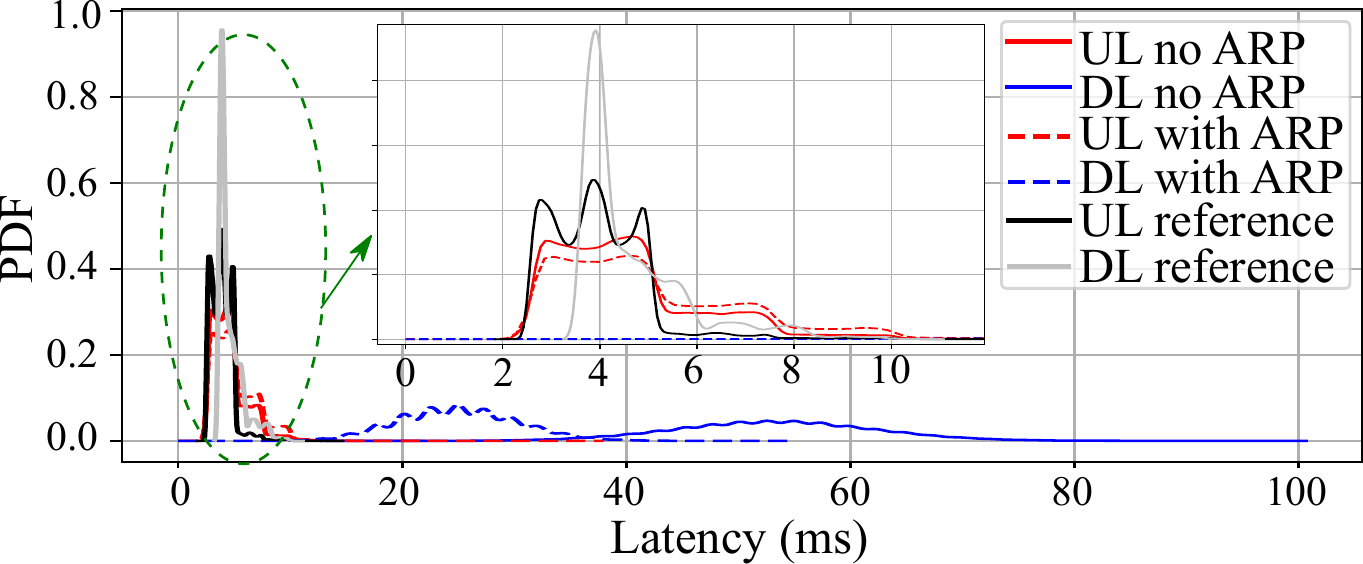}
	}
	\caption{The impact of QoS ARP on the UL and DL one-way \ac{OTA} latency performance in a scenario with 2 UEs with maximum additional DL traffic load on both UEs.}
	\label{fig:latency_2_UEs}
\end{figure}
\begin{figure}[t!]
	\centering
	\subfloat[CCDF.\label{fig:UL_CCDF_2_UEs_QoS_no_QoS}]{%
		\includegraphics[width=0.48\textwidth]{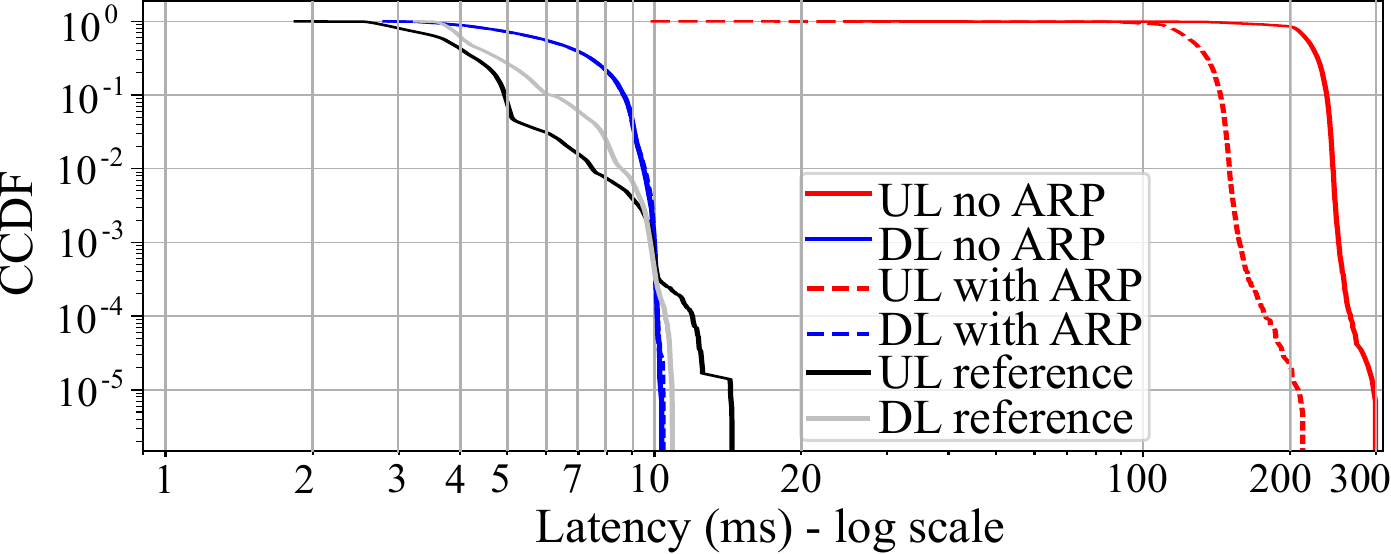}
	}\\
	\subfloat[PDF.\label{fig:UL_PDF_2_UEs_QoS_no_QoS}]{%
		\includegraphics[width=0.48\textwidth]{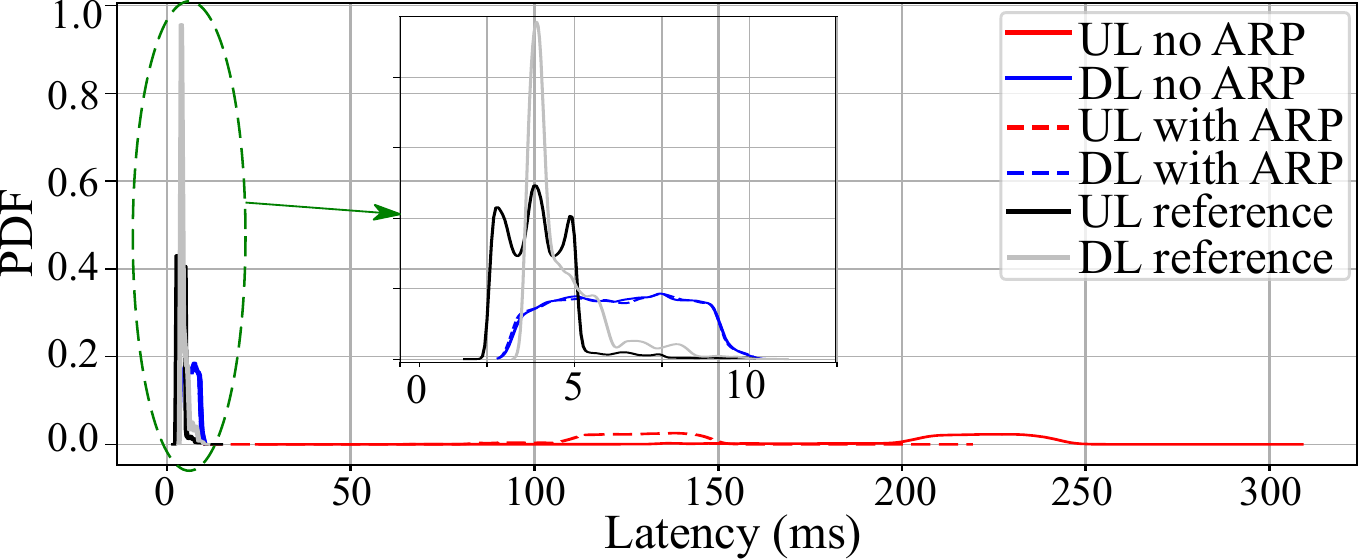}
	}
	\caption{The impact of QoS ARP on the UL and DL one-way \ac{OTA} latency performance in a scenario with 2 UEs with maximum additional UL traffic on both UEs.}
	\label{fig:UL_latency_2_UEs}
        \vspace{-3mm}
\end{figure}
In this section, we provide the \ac{E2E} latency measurement results with varying channel conditions of a UE, the impact of the QoS ARP prioritization on the UL and DL latency, and the impact of a high traffic load on the latency. In all scenarios, latency is measured with \textit{IRTT} tool, generating random-value packets of 60$\,$Bytes and cyclic interval of 10$\,$ms for the duration of an hour (360000 packets).
 
In the first scenario, the impact of channel conditions on the one-way latency with fixed additional DL traffic load (generated using \textit{iperf3} tool) is presented in Fig.$\,$\ref{fig:latency_good_bad}. It can be seen that the DL latency increases when channel conditions are poor due to higher resource occupation with decreased \ac{PDSCH} \ac{MCS} for a fixed DL traffic load of 450$\,$Mbps. On the other hand, the traffic load in DL slightly affects the UL latency, comparing scenarios of good and poor channel conditions.

The second scenario shows the importance of the QoS ARP parameter in terms of latency if the maximum DL traffic load is generated for multiple UEs. The mean value of DL latency decreases from 53$\,$ms to 25$\,$ms when using QoS ARP prioritization whereby the measured DL latency in the reference scenario without additional traffic load is 4.6$\,$ms. Moreover, it can be seen from Fig.$\,$\ref{fig:latency_2_UEs} that the UL latency also increases slightly when the DL traffic load is high. This behavior is specifically noticeable at the tail of the latency distribution curve compared with the reference scenario. In these scenarios, the latency is measured at the UE with a higher QoS ARP level.

Similar to the previous scenario, we also tested a scenario where the maximum UL traffic load was generated by multiple UEs, utilizing all the available UL resources and therefore causing congestion in the UL traffic. Fig.$\,$\ref{fig:UL_latency_2_UEs} shows the impact of the QoS ARP parameter on latency performance. The mean value of UL latency decreases from 211$\,$ms to 123$\,$ms, while the mean value measured for the reference scenario without additional traffic load is 3.9$\,$ms. From both Fig.$\,$\ref{fig:UL_CCDF_2_UEs_QoS_no_QoS} and \ref{fig:UL_PDF_2_UEs_QoS_no_QoS} it can be seen that there is no impact of QoS ARP parameter on the DL latency, in scenarios with maximum UL traffic load. However, compared with the reference scenario  without additional traffic load, there is also a noticeable increase in DL latency, in scenarios with maximum UL traffic load. It is important to note that only in scenarios with maximum UL traffic load, a significant packet loss of UL latency-measurement packets was observed, i.e., 29\% with QoS ARP and 52\% without QoS ARP.

From the measurements results, it can be concluded that the E2E OTA latency with low traffic load, as shown in Fig. \ref{fig:latency_good_bad}-\ref{fig:UL_latency_2_UEs} as \lq UL reference\rq and \lq DL reference\rq has a mean value of 3.9$\,$ms and 4.6$\,$ms, and a maximum value of 14$\,$ms and 11$\,$ms, respectively, with the reliability of 99.9997$\,$\%, based on the measurement duration of an hour. This performance could fulfill requirements of several industrial use cases, such as monitoring or remote maintenance, as defined in Table 1 in \cite{9299391}, industrial wireless sensors use cases, and mobile robots (real-time video stream) use case, defined by 3GPP in \cite{standard:service_req_3gpp} in Table 5.2-2 and Table 5.4-1, respectively. However, with a very high traffic load, the latency values increase significantly. The QoS ARP level improves the latency performance, however, it is not sufficient to ensure the controlled latency performance. First, 5G \ac{QoS} parameters such as  \ac{MFBR} or \ac{AMBR} could be used to prevent traffic congestion which can significantly decrease the maximum latency values. Second, 5G QoS parameters such as \ac{5QI}, the ARP-\textit{preemption capability} option, or \ac{URLLC} features are necessary in order to control the maximum latency in all channel or traffic load conditions.

\section{Conclusions}
\label{sec:conclusion}
The use of 5G for several different vertical applications is highly advocated. In this work, we evaluate the performance of the 5G campus network considering factory floor scenarios. It can be concluded from the coverage heatmaps that the placement of the \acp{RRU} is suitable to achieve high communication performance in the factory floor area. The TP results show a minimum of 450$\,$Mbps in DL and 120$\,$Mbps in UL is available in the factory area. These results can serve as a guide in planning for the placement of RRUs in indoor environments. It can be concluded from the TP measurements that with the QoS ARP parameter, the share of available resources among different QoS flows can be controlled in a deterministic manner, even in scenarios with extremely high traffic load as well as in various channel conditions. However, the channel conditions have to be taken into account while planning the required resource share for a certain QoS flow such that the required TP can be achieved.

On the contrary, from the latency measurement results, it can be concluded that latency increases significantly with high traffic load and cannot be controlled in a deterministic manner using only the QoS ARP levels, even though it was observed that the latency is noticeably lower for high-priority QoS flows. Moreover, it can be concluded that variations in latency for the same traffic load depend on channel conditions, due to the utilization of a different amount of resources with a varying modulation scheme.

Lastly, the presented measurement results show that this 5G setup could fulfill several industrial use cases, defined by 3GPP or 5G ACIA in terms of TP and latency performance, even without the usage of \ac{URLLC} features. However, latency performance improvements are expected with \ac{URLLC} features.
\section*{Acknowledgment}
This work is funded under the research and technology development of gigabit applications as part of lighthouse projects BBA2030:GA of the Federal Ministry for Finance, Austria represented by the Austrian Research Promotion Agency (FFG) under the grant agreement no. FO999899772.
\bibliographystyle{IEEEtran}
\bibliography{IEEEabrv,IECON}

\end{document}

%% file: acronyms.tex
\begin{acronym}
	
	\acro{3GPP}[3GPP]{3\textsuperscript{rd} generation partnership project}
        \acro{5QI}[5QI]{5G \ac{QoS} identifier}
	\acro{API}[API]{application programming interface}
        \acro{ACIA}[ACIA]{alliance for connected industries and automation}
	\acro{AUSF}[AUSF]{authentication server function}
	\acro{AMF}[AMF]{access and mobility management function}
        \acro{AMBR}[AMBR]{aggregated maximum bit rate}
	\acro{ARP}[ARP]{allocation and retention priority}
	\acro{BBU}[BBU]{baseband unit}
	\acro{BW}[BW]{bandwidth}
	\acro{CP}[CP]{control plane}
	\acro{CQI}[CQI]{channel quality indicator}
	\acro{CSI}[CSI]{channel state information}
	\acro{DL}[DL]{downlink}
	\acro{E2E}[E2E]{end to end}
	\acro{gNB}[gNB]{next generation base station}
        \acro{GFBR}[GFBR]{guaranteed flow bit rate}
	\acro{5GS}[5GS]{5G system}
	\acro{IRTT}[IRTT]{isochronous round-trip tester}
        \acro{iperf}[iperf]{internet performance working group}
	\acro{JKU}[JKU]{Johannes Kepler University, Linz}
	\acro{LIT}[LIT]{Linz institute of technology}
	\acro{MCS}[MCS]{modulation and coding scheme}
        \acro{MFBR}[MFBR]{maximum flow bit rate}
        \acro{MIMO}[MIMO]{multiple input multiple output}
	\acro{NIC}[NIC]{network interface card}
	\acro{NRF}[NRF]{network repository function}
	\acro{OS}[OS]{operating system}
	\acro{OIC}[OIC]{open innovation center}
        \acro{OTA}[OTA]{over-the-air}
	\acro{PTP}[PTP]{precision time protocol}
	\acro{ptp4l}[ptp4l]{\ac{PTP} for Linux}
	\acro{PDSCH}[PDSCH]{physical downlink shared channel}
	\acro{QoS}[QoS]{quality of service}
        \acro{QAM}[QAM]{quadrature amplitude modulation}
	\acro{RAN}[RAN]{radio access network}
	\acro{RRU}[RRU]{remote radio unit}
	\acro{RHUB}[RHUB]{\ac{RRU} HUB}
	\acro{RSRP}[RSRP]{reference signal received power}
	\acro{RTT}[RTT]{round-trip time}
	\acro{RX}[RX]{receive}
	\acro{RB}[RB]{resource block}
	\acro{SCS}[SCS]{subcarrier spacing}
	\acro{SA}[SA]{standalone}
	\acro{SMF}[SMF]{session management function}
	\acro{UP}[UP]{user plane}
	\acro{UE}[UE]{user equipment}
	\acro{UL}[UL]{uplink}
	\acro{UDM}[UDM]{unified data management function}
	\acro{UPF}[UPF]{user plane function}
        \acro{URLLC}[URLLC]{ultra-reliable low-latency communication}
	\acro{TP}[TP]{throughput}
	\acro{TX}[TX]{transmit}
        \acro{TCP}[TCP]{transmission control protocol}
	\acro{TDD}[TDD]{time division duplex}
        \acro{IoT}{internet of things}

\end{acronym}